\documentstyle[12pt]{article}
\begin{document}
\begin{center}
{\bf  QUANTUM INTERFERENCE AND MONTE-CARLO
 SIMULATIONS OF MULTIPARTICLE PRODUCTION}\\
\vspace {1cm}
A. Bialas\footnote{Permanent address: 
Institute of  Theoretical Physics, Jagellonian
 University, Cracow, Poland} and A. Krzywicki\\
Laboratoire de Physique Th\'{e}orique et Hautes 
Energies\\ B\^{a}t. 211, Universit\'{e} de 
Paris-Sud, 91440 Orsay, 
France\footnote{Laboratoire
 associ\'e au C.N.R.S., URA-D0063}\\
\end{center}
\vspace {2cm}
{\bf Abstract} : 
We show that the effects of quantum 
interference can be implemented in 
Monte-Carlo generators by
modelling the generalized Wigner 
functions. A specific prescription 
for an appropriate modification of
the weights of events produced by 
standard generators is proposed.

\vspace {3cm}
\par\noindent
April 1995\\
LPTHE Orsay 95/31    \\
hep-ph/9504349 \\
\newpage
\par
    {\bf  1}. A quantitative analysis of multiparticle 
data requires using generators of computer
"events". The trouble is that the models
 employed in this context
are essentially classical and that their
 predictions are inadequate and sometimes turn out
to be misleading in the study of those aspects of 
multiparticle production where quantum 
interference is important \cite{cfk}. This 
is a serious problem because, as evidenced 
by the studies of the so-called intermittency
 phenomenon \cite{bp}, the momentum-space 
short-range correlations are to a large extent
 dominated by the Hanbury-Brown-Twiss (HBT) 
effect \cite{HBT}, i.e. by quantum interference.
\par
We would like to emphasize that this does not
mean at all that these correlations are "trivial" 
as is sometimes claimed. On the 
contrary, since quantum interference is 
sensitive to 
the space-time development of the 
collision process it yields precious
 information thereon and its effects
 should be examined with most attention.
\par
The aim of this communication is to outline 
a way of implementing the effects of 
quantum statistics in 
Monte-Carlo simulations of multiparticle
phenomena. One could wonder whether this 
is possible at
all, since the Monte-Carlo method deals with 
probabilities while quantum interference is
a consequence of adding amplitudes whose
 phases are essential. We will show, however,
 that one can devise a systematic and 
practical approach to the problem if 
Wigner functions and not scattering 
amplitudes are used to describe 
the multiparticle system.
\par
{\bf 2}. We shall now write the n-particle 
spectrum in terms of the Wigner 
function, first for distinguishable 
and then for identical secondaries.
\par
Let $\psi(\mbox{{\bf q}}; \alpha)$ 
denote a wave-function describing a
 stationary state produced at high-energy.
 Here $\mbox{{\bf q}}$ refers to the
 momenta $q_1, q_2, ...,q_ n$ of the $n$ 
produced spinless particles and $\alpha$ 
denotes all other parameters, assumed 
irrelevant for our problem. For the
 moment, the secondaries are supposed
 to be distinguishable.
\par
The n-particle spectrum is, of course

\begin{equation}
\Omega_0(\mbox{{\bf q}}) = \sum_{\alpha}
 \mid \psi(\mbox{{\bf q}}; \alpha) \mid ^2 
, \; \; \; \mbox{{\bf q}} = (q_1 , ..., q_n)
\label{1}
\end{equation}

\noindent
The weights generated by a Monte-Carlo 
algorithm are directly  proportional to
  $\Omega_0(\mbox{{\bf q}})$.
\par
Going over to the coordinate
 representation we write

\begin{equation}
\Omega_0(\mbox{{\bf q}}) = \int
 d \mbox{{\bf x}} d \mbox{{\bf x}}' 
e^{i \mbox{{\bf q}} \cdot (\mbox{{\bf x}}
 - \mbox{{\bf x}}')}  \; \rho 
(\mbox{{\bf x}}; \mbox{{\bf x}}')
 , \; \; \; \mbox{{\bf x}} = (x_1 , ... , x_n)
\label{2}
\end{equation}

\noindent
where

\begin{equation}
\rho(\mbox{{\bf x}}; 
\mbox{{\bf x}}') = \sum_{\alpha} 
\hat{\psi}(\mbox{{\bf x}};
 \alpha) \hat{\psi}^*
 (\mbox{{\bf x}}' ; \alpha)
\label{3}
\end{equation}

\noindent
is a density matrix, $\hat{\psi}$ 
being the Fourier transform of 
$\psi$. Eq. (\ref{2}) can further
 be rewritten as

\begin{equation}
\Omega_0(\mbox{{\bf q}}) = \int 
d\mbox{{\bf x}}^{+}  \; W(\mbox{{\bf q }};
 \mbox{{\bf x}}^{+})
\label{4}
\end{equation}

\noindent
where [notation: $\mbox{{\bf x}}^{+}
 ={1 \over 2} (\mbox{{\bf x + x}}'),
 \mbox{{\bf x}}^{-} = \mbox{{\bf x - x}}'$]

\begin{equation}
W(\mbox{{\bf q}} ; \mbox{{\bf x}}^{+})
 = \int d\mbox{{\bf x}}^{-} e^{i
 \mbox{{\bf q}} \cdot 
\mbox{{\bf x}}^{-}} \rho(\mbox{{\bf x}}
 ; \mbox{{\bf x}}')
\label{5}
\end{equation}

\noindent
is the generalized Wigner function 
\cite{wig}, the quantum analog of 
the classical Boltzmann phase-space
 density. It is real and it gives 
the observable spectrum when 
integrated, as seen in (\ref{4}).
\par
Let us now assume that the secondaries
 are all identical
(the generalisation of the discussion 
to the case 
where there are several species of 
identical secondaries 
is straightforward). Let $\cal{P}$ 
denote some arbitrary permutations
 of the integers $1, ..., n$ and 
$\mbox{{\bf q}}_{\cal{P}}$ the
 corresponding permutation of the 
momenta $q_1, ..., q_n$.  Once the
 wave function 
$\psi (\mbox{{\bf q}} ; \alpha)$ 
is symmetrized with respect to the
 momenta of produced bosons, eq. 
(\ref{2}) becomes

\begin{equation}
\Omega (\mbox{{\bf q}}) = {1 \over {n!}} 
\sum_{\cal{P},\cal{P'}}  \; \int 
d\mbox{{\bf x}} d\mbox{{\bf x}}' 
 e^{i (\mbox{{\bf x}} \cdot 
\mbox{{\bf q}}_{\cal{P}} -  \mbox{{\bf x}}' 
\cdot \mbox{{\bf q}}_{\cal{P'}})} \; 
\rho (\mbox{{\bf x}} ; \mbox{{\bf x}}')
\label{6}
\end{equation}

\noindent
which is rewritten as

\begin{equation}
\Omega(\mbox{{\bf q}}) = {1 \over {n!}} 
\sum_{\cal{P},\cal{P}'} \;
 \int d \mbox{{\bf x}}^{+} 
 e^{i \mbox{{\bf x}}^{+} \cdot 
(\mbox{{\bf q}}_{\cal{P}} - 
\mbox{{\bf q}}_{\cal{P}'})} \; 
W({{\mbox{{\bf q}}_{\cal{P}} + 
\mbox{{\bf q}}_{\cal{P}'}} \over 2} ; 
\mbox{{\bf x}}^{+})
\label{7}
\end{equation}

\noindent
with $W(\mbox{{\bf q}}; \mbox{{\bf 
x}}^{+})$ defined in (\ref{5}). 
Thus we observe that the same Wigner
 function determines the spectrum
 before and after the symmetrization.
 This has a simple physical reason:
 All information, compatible with 
the rules of quantum mechanics, 
about what happens in the full 
phase-space of coordinates and 
momenta (in contrast to the 
momentum-space alone) is encoded 
in the Wigner function. This 
information includes the 
phases of different waves and
 therefore all that is 
needed to predict the interference
 patterns. A further advantage of 
the Wigner function is that it 
appeals directly to one's 
intuition. A few words of caution
 are necessary at this point, however.
\par
Contrary to the Boltzmann phase-space
 density, the Wigner function is 
locally not positive definite. It
 can actually oscillate quite violently. 
The oscillations integrate to zero 
in (\ref{5}) but can conspire with 
oscillating terms in the integrand 
of (\ref{7}) to contribute 
significantly to the result. This
 is how quantum mechanics shows up
 in the problem. Thus, it is clear
 that regarding the Wigner function
 as a phase-space density is possible
 only when the function is smoothed 
by averaging the oscillations out. 
Physically it means appropriate
 smearing of coordinates and momenta. 
The price to pay is that, in general,
 the resulting
 probabilistic model can only be 
trusted when the momentum differences
 appearing in (\ref{7}) are not too large. 
\par
{\bf 3}. The standard Monte-Carlo 
algorithms rest on models 
of momentum-space densities. The 
goal to achieve is to correct the 
weights of Monte-Carlo events, once 
they have been
generated according to the 
distribution $\Omega_0(\mbox{{\bf q}})$.
 Our proposal consists in going from 
$\Omega_0(\mbox{{\bf q}})$ to
 $\Omega(\mbox{{\bf q}})$ by 
modelling the Wigner function. 
\par
Writing
\begin{equation}
W(\mbox{{\bf q}} ; 
\mbox{{\bf x}}^{+}) = 
\Omega_0(\mbox{{\bf q}}) \;
 w(\mbox{{\bf q}}; \mbox{{\bf x}}^{+}) \; ,
\label{8a}
\end{equation}

\noindent
we see that, if the Wigner function is 
regarded as a phase-space density, $w(\mbox{{\bf q}};
 \mbox{{\bf x}}^{+})$ has the
 meaning of a conditional probability:
 given that the particles with 
momenta $q_1, ... q_n$ are present 
in the final state, $w$ is the 
probability that they are 
produced at the points
 $x_1^{+},..., x_n^{+}$. 
The problem is to 
construct a viable model for 
$w(\mbox{{\bf q}}; \mbox{{\bf x}}^{+})$.
\par
In the absence of additional 
information it seems reasonable
 to start with the working 
assumption that the likelihood
 to radiate a particle from a 
given space point is statistically
 independent of what happens to
 other particles. This means
 that $w$ factorizes:

\begin{equation}
w(\mbox{{\bf q}}; \mbox{{\bf 
x}}^{+}) = \prod_{j} w(q_j,x_j^{+})
\label{8b}
\end{equation}

\noindent
where

\begin{equation}
\int d^3x \; w(q,x) = 1
\label{9}
\end{equation}

\noindent
Substituting (\ref{8a})-(\ref{8b})
 into (\ref{7}) one finds

\begin{equation}
\Omega (\mbox{{\bf q}}) = 
{1 \over  {n!}} 
\sum_{\cal{P},\cal{P'}} 
\Omega_0({\mbox{{\bf q}}_{\cal{P}}
 + \mbox{{\bf q}}_{\cal{P'}} 
\over 2} ) \prod_{j} 
\hat{w} [q_j, (q_{\cal{P}} -
 q_{\cal{P'}})_j]
\label{10}
\end{equation}

\noindent
where

\begin{equation}
\hat{w}(q,\Delta) = \int d^3x \; 
e^{ix \cdot \Delta} w(q,x)
\label{11}
\end{equation}

\noindent
Eq. (\ref{9}) and the re\-a\-li\-ty
 of $w(q,x)$ im\-ply $\hat{w}(q,0) 
= 1$ and $\hat{w}(q, \Delta) = 
\hat{w}^* (q, - \Delta)$, 
res\-pec\-ti\-ve\-ly. This 
gua\-ran\-tees that 
$\Omega(\mbox{{\bf q}})$ 
cal\-cu\-la\-ted from (\ref{10}) is real.
\par
Eq. (\ref{10}) can be used as 
it stands when one has an 
explicit formula for 
$\Omega_0(\mbox{{\bf q}})$. 
In practice, however, 
$\Omega_0(\mbox{{\bf q}})$ is 
constructed iteratively by a 
Monte-Carlo algorithm and at a given
stage of the simulation it is 
computed for one configuration 
of momenta. To deal with
 this complication we observe 
that one does not make a big 
error by replacing in 
(\ref{10}) $\Omega_0({\mbox{{\bf q}}_{\cal{P}}
 + \mbox{{\bf q}}_{\cal{P'}} \over 2})$ 
by $\Omega_0(\mbox{{\bf q}}_{\cal{P}})$. 
Indeed, those terms in eq. (\ref{10}) where 
this approximation is poor are
 suppressed by the rapidly 
decreasing factors $\hat{w}$ and thus need not 
 be calculated with a great precision. The equation
 (\ref{10}) now becomes

\begin{equation}
\Omega (\mbox{{\bf q}}) = {1 \over
  {n!}} \sum_{\cal{P}} \{ 
\Omega_0(\mbox{{\bf q}}_{\cal{P}}) 
\; \sum_{\cal{P}'} \; \prod_{j} 
\hat{w} [q_j, (q_{\cal{P}}
 - q_{\cal{P'}})_j] \}
\label{10bis}
\end{equation}

\noindent
Thus, once a configuration 
$\mbox{{\bf q}}_{\cal{P}}$ of momenta
 has been generated by the original 
algorithm, the weight of the event 
in question has to be multiplied by 
the correction factor $\mbox{{\rm Re}}\{ \sum_{\cal{P}'} 
\; \prod_{j} 
\hat{w} [q_j, (q_{\cal{P}}
 - q_{\cal{P'}})_j]\}$ in order to 
take care of the HBT interference.
 This is the result sought. In (\ref{10bis})
the sum over $\cal{P}$ 
just expresses formally the fact that
 even in a classical model 
the labelling of identical 
particles is merely a matter of convention.
\par
The function $\hat{w}(q,\Delta)$
 is unknown. It can either
 be taken from a model \cite{models}
 or, perhaps more reliably, be 
determined  by fitting 2-body HBT 
correlations. The weight of an 
event is then completely determined 
and the procedure can be used to
 study the implications of quantum
 statistics for other aspects 
of the production process.
\par
{\bf 4}. Adopting the probabilistic
 interpretation of the Wigner
 function, the proposed approach allows 
an intuitive interpretation of the
 results in terms of the 
space-time structure 
of the region of particle emission.
 As already mentioned 
this is meaningful when momentum
 differences are not too large. 
\par
 It should be clear that a simple 
physical meaning can be ascribed 
to $w(q, x)$ rather than to $\hat{w}(q,
 \Delta)$. Let us briefly outline 
how one might proceed in modelling 
$w(q, x)$. We imagine that a 
particle with momentum $q$ is
 emitted from a diffuse source 
centered at $x = x_0(q)$. We 
are thus led to write

\begin{equation}
w(q, x) = P[q, x - x_0(q)]
\label{12}
\end{equation}

\noindent
The simplest choice for $P(q, x)$ 
would be to take a Gaussian

\begin{equation}
P_G(x) = {1 \over {\pi^{3 
\over 2}\sigma_x \sigma_y \sigma_z 
  }} e^{-(x_x^2/\sigma_x^2 + 
x_y^2/\sigma_y^2 + x_z^2/\sigma_z^2)}
\label{14}
\end{equation}

\noindent
with $\sigma = \sigma(q)$, to 
take into account the possible
 dependence of the shape of the
 source on $q$. More generally one can set

\begin{equation}
P(q, x) = \int d^3\sigma \; 
H(q, \sigma) P_G(x)
\label{15}
\end{equation}

This gives

\begin{equation}
\hat{w}(q, \Delta) = e^{i 
\Delta \cdot x_0(q)} 
\int d^3\sigma \; H(q, 
\sigma) \; e^{-{1
 \over 4}(\sigma_x^2 \Delta_x^2 + 
\sigma_y^2 \Delta_y^2 
+ \sigma_z^2 \Delta_z^2)} ,
\label{16}
\end{equation}

\noindent
a formula that seems general 
enough to accomodate all 
physically reasonable choices for $\hat{w}$.
\par
Notice that, for obvious physical
 reasons, the distribution (\ref{10})
 can only depend on differences 
$x_0(q) - x_0(q')$. This is indeed 
the case as can be seen by 
observing that $\sum_{j} (q_{\cal{P}} 
- q_{\cal{P}'})_j = 0 $ for any 
two permutations $\cal{P}$ and $\cal{P}'$. 
\par
A further simplification is 
obtained by assuming 
that the place where a particle
 is produced depends at most on 
some global characteristics of 
the collision, like the total 
energy, to give an example. This
 is presumably a good assumption
 as long as the source can be 
regarded as static. Technically,
 it amounts to neglect the
 first argument of 
$\hat{w}(q,\Delta)$. 
\par
{\bf 5}. A few comments are in order:
\par
(i) The proposed approach,
 even used in its simplest 
version, enables one to incorporate into a 
Monte-Carlo simulation the {\em collective nature} 
of the HBT effect.
\par
(ii) Our ansatz (\ref{8b}) is to be 
considered as a 
working assumption to be
 upgraded when more information
is available (e.g. if two
 identical particles result 
from the decay of the same resonance).
\par
(iii) The ansatz can be
 checked against the data on higher 
order correlations. This may lead to
 a discovery of hitherto entirely 
unknown correlation structures in 
the space-time development of the 
collision and is, therefore, 
potentially of great interest. 
\par
(iv) The motivation of our
 probabilistic approach to
 the Wigner function is
 mostly phenomenological. It would be, 
of course, highly desirable 
to invest more effort in 
developing theoretically 
based models of the Wigner 
function.
\par
To summarize, we have shown that
 the effects of quantum interference
 in multiparticle production can be 
naturally expressed in terms of 
generalized Wigner functions. Such 
formulation allows one to incorporate 
these effets into Monte-Carlo 
generators and to give them an 
intuitive interpretation in terms 
of the space-time development of 
the interaction. It also explicitly 
demonstrates that in order to obtain
 theoretically founded predictions 
for the HBT correlations one should
 compute the Wigner functions from
 the underlying theory.
\vspace{1cm}
\par
{\bf Acknowledgements}: We would 
like to thank J.-P. Blaizot for
 helpful remarks and W.J. Metzger
for an illuminating correspondence. 
This work was partly supported by 
the KBN grant N$^{\rm o}$ 2 PO3B 08308.


\begin{thebibliography}{99}
\bibitem{cfk} See for example
 A. Capella , K. Fia\l kowski 
and A. Krzywicki, Physics
 Letters B230 (1989) 149 .
\bibitem{bp} A. Bialas and 
R. Peschanski, Nucl. Phys. 
B273 (1986) 703   and B308 
(1988) 857 ; for recent reviews 
see: P. Bozek, M. Ploszajczak 
and R. Botet, Phys. Rep. 252 
(1995) 101; E.A. De Wolf, I.M.
 Dremin and W. Kittel, 
preprint HEN-363 (1993).
\bibitem{HBT} R. Hanbury-Brown 
and R.Q. Twiss, Nature, 177 (1957) 
27. For a review see D.H.
 Boal, C.K. Gelbke
 and B.K. Jennings, Rev. Mod.
 Phys. 62 (1990) 553.
\bibitem{wig} P. Carruthers and
 F. Zachariasen, Rev. Mod. Phys.
 55 (1983) 245 and references therein; 
S. Pratt, Phys. Rev. Lett. 53 (1984) 1219.
\bibitem{models} Several models 
for $\hat{w}$ have been discussed. 
See e.g. S. Pratt, T. Csorgo 
and J. Zymanyi, Phys. Rev. C42 
(1990) 2646; I.V. Andreev, M. 
Plumer and R.M. Weiner, Int.
 J. Mod. Phys. A8 (1993) 4577 
and references therein.
 For a review see \cite{HBT}.
\end{thebibliography}
\end{document}